\documentclass{article}
\usepackage{amssymb}
\usepackage{fleqn}
\usepackage{epsfig}
\usepackage{graphicx}
\usepackage{amsmath}
\def\beq{\begin{equation}}
\def\eeq{\end{equation}}
\def\bea{\begin{eqnarray}}
\def\eea{\end{eqnarray}}
\def\bq{\begin{quote}}
\def\eq{\end{quote}}

\def\gappeq{\mathrel{\rlap {\raise.5ex\hbox{$>$}}
{\lower.5ex\hbox{$\sim$}}}}
\def\lappeq{\mathrel{\rlap{\raise.5ex\hbox{$<$}}
{\lower.5ex\hbox{$\sim$}}}}
\def\Toprel#1\over#2{\mathrel{\mathop{#2}\limits^{#1}}}

\begin{document}

\pagestyle{empty} 
\begin{flushright}
DSF-4/2002
\end{flushright}
\vspace*{15mm}

\begin{center}
\textbf{APPROXIMATE NNLO THRESHOLD RESUMMATION  
IN HEAVY FLAVOUR DECAYS } \\[0pt]

\vspace*{1cm}

\textbf{U. Aglietti} \\[0pt]

\vspace{0.3cm} CERN-TH, Geneva, Switzerland,\\
Dipartimento di Scienze Fisiche,\\ 
Universit\'a di Roma ``La Sapienza'', \\
and I.N.F.N.,
Sezione di Roma, Italy. \\
e-mail address: ugo.aglietti@cern.ch.\\[1pt]

\vspace{0.3cm}\textbf{G. Ricciardi} \\[0pt]

\vspace{0.3cm} Dipartimento di Scienze Fisiche,\\ 
Universit\'a di Napoli ``Federico II'' \\
and I.N.F.N.,
Sezione di Napoli, Italy. \\
e-mail address: giulia.ricciardi@na.infn.it\\[1pt]

$~~~$ \\[0pt]
\vspace*{2cm} \textbf{Abstract} \\[0pt]
\end{center}

We present an approximate
 NNLO evaluation of the QCD form factor resumming large
logarithmic perturbative contributions in semi-inclusive heavy flavour decays.
\vspace*{6cm} \noindent 

\vfill\eject

\setcounter{page}{1} \pagestyle{plain}

\section{Introduction}

In this note we present an approximate
 next-to-next-to-leading order (NNLO) evaluation of
the QCD form factor resumming threshold logarithmic contributions in
semi-inclusive heavy flavour decays.

In semi-inclusive processes, final gluon radiation is strongly inhibited in the
phase space regions where the observed final state obtains its maximum energy,
therefore
opening the way to soft and collinear singularities.
The perturbative calculation of the differential cross section is
plagued, in that limit, by large logarithms. In order to improve the
 reliability of the perturbative calculation, these large logarithms
need to be resummed.

Let us consider the rate of  semi-inclusive
 heavy flavour decays.
The Mellin $N$-moments of the rate contain double logarithmic
contributions and have an expansion of the form: 
\begin{eqnarray}
\frac{1}{\Gamma _{B}}\Gamma _{N}\left( \alpha _{S}\right) 
&\equiv& \int_{0}^{1}dx\,x^{N-1} \,
 \frac{1}{\Gamma _{B}} \, \frac{d\Gamma }{dx}\left( x;\alpha _{S}\right) \\
&=&1+\sum_{n=1}^{\infty }\sum_{m=0}^{2n}G_{n\,m}\,\alpha
_{S}^{n}\,L^{m}+\sum_{l=1}^{\infty }\alpha _{S}^{l}\,R_{N}^{\left( l\right) }
\notag \\
&=&1+G_{12}\,\alpha _{S}\,L^{2}+G_{11}\,\alpha _{S}\,L+G_{10}\,\alpha
_{S}\,+G_{24}\,\alpha _{S}^{2}\,L^{4}+G_{23}\,\alpha
_{S}^{2}\,L^{3}+G_{22}\,\alpha _{S}^{2}L^{2}+\,\cdots   \notag \\
&&\,\,\,\,+\alpha _{S}\,R_{N}^{\left( 1\right) }+\alpha
_{S}^{2}\,R_{N}^{\left( 2\right) }+\alpha _{S}^{3}\,R_{N}^{\left( 3\right)
}+\cdots ,
\label{exp}
\end{eqnarray}
where 
\begin{equation}
L\equiv \ln N.
\end{equation}
$R_{N}\left( \alpha _{S}\right) $ is a remainder function, which does not
contain large logarithms and has a perturbative expansion of the form: 
\begin{equation}
R_{N}\left( \alpha _{S}\right) \equiv \sum_{l=1}^{\infty }\alpha
_{S}^{l}\,R_{N}^{\left( l\right) }.
\end{equation}
The running coupling is evaluated at a general renormalization scale $\mu
\neq Q,$ where $Q$ is the hard scale: 
\begin{equation}
\alpha _{S}=\alpha _{S}\left( \mu ^{2}\right) .
\end{equation}
The logarithmic terms have an exponential structure, so one can write \cite
{cattren,cattren2}: 
\begin{equation}
\frac{1}{\Gamma _{B}}\Gamma _{N}\left( \alpha _{S}\right) \,=\,C\left(
\alpha _{S}\right) \,\,f_{N}\left( \alpha _{S}\right)   \label{quasifinale}
\end{equation}
where the form factor $f_{N}\left( \alpha _{S}\right) $ reads: 
\begin{eqnarray}
\,f_{N}\left( \alpha _{S}\right)  &=&\exp \left[ \sum_{n=1}^{\infty
}\sum_{k=1}^{n+1}c_{n\,m}\,\alpha _{S}^{n}\,L^{k}\right]   \notag \\
&=&\exp \left[ c_{12}\alpha _{S}\,L^{2}+c_{11}\alpha _{S}\,L+c_{23}\alpha
_{S}^{2}\,L^{3}+c_{22}\alpha _{S}^{2}\,L^{2}+c_{2\,1}\alpha
_{S}^{2}\,L+c_{34}\alpha _{S}^{3}\,L^{4}+\cdots \right] .
\end{eqnarray}
Note that the exponent contains only the first term $\alpha _{S}\,L^{2}$ of
the double-logarithmic series $\left( \alpha _{S}\,L^{2}\right) ^{n}.$
The advantage of the exponentiation is
therefore that we can  predict $\ln f_N$ reliably for $\alpha_S 
L \ll 1$, that is, for a larger region than  $\alpha_S 
L^2 \ll 1$, where the perturbative expansion of $f_N$ holds.
In fact, the other terms of $(\alpha_S L^2)^n$ come purely from the expansion of the
 exponential function, as can be seen in formula (\ref{exp}).

 The
prefactor in (\ref{quasifinale}) is the coefficient function, having an
expansion in powers of $\alpha_{S}$: 
\begin{equation}
C\left( \alpha _{S}\right) =1+\sum_{n=1}^{\infty }C_{n}\,\alpha
_{S}^{n}=1+C_{1}\alpha _{S}+C_{2}\alpha _{S}^{2}+\cdots .
\end{equation}
The double sum in the exponent is usually organized as a series of
functions, which resum ``strips'' in the $\left( n,k\right) $ plane: 
\begin{eqnarray}
f_{N}\left( \alpha _{S}\right)  &=&\exp \left[ L\,g_{1}\left( \beta
_{0}\alpha _{S}L\right) +\sum_{n=2}^{\infty }\alpha _{S}^{n-2}\,g_{n}\left(
\beta _{0}\alpha _{S}L\right) \right]   \notag \\
&=&\exp \left[ L\,g_{1}\left( \beta _{0}\alpha _{S}L\right) +\,g_{2}\left(
\beta _{0}\alpha _{S}L\right) +\alpha _{S}\,g_{3}\left( \beta _{0}\alpha
_{S}L\right) +\alpha _{S}^{2}\,g_{4}\left( \beta _{0}\alpha _{S}L\right)
+\cdots \right].
\end{eqnarray}
The functions $g_{i}\left( \lambda \right) $ have a power-series expansion: 
\begin{equation}
g_{i}\left( \lambda \right) =\sum_{n=1}^{\infty }g_{in}\lambda ^{n}.
\end{equation}
where $\lambda=\beta_0 \alpha_S L$.
They are all homogeneous functions: $g_{i}\left( 0\right) =0.$ This property
insures the normalization of the form factor: $f_{N=1}=1.$ 
The resumming at LO, NLO and NNLO is referred, respectively,
 to series in $ L (\alpha_S L)^n$, $ (\alpha_S L)^n$ and
$ \alpha_S (\alpha_S L)^n$.
  
We have computed the function $g_3(\lambda)$, which is necessary for the NNLO
resumming.
Not all the quantities determining $g_3(\lambda)$ are 
 known exactly. $A_{3}$ is only known numerically from a fit of the
three-loop Altarelli-Parisi (AP) splitting function to the known moments. $%
B_{2}$ is unknown and we approximate it with the infrared-regular part of
the two-loop AP splitting function.
 The coefficient functions and the remainder
functions for radiative and semileptonic $B$ decays
 have been computed to NLO in
ref. \cite{menpb}. 
A complete NNLO computation of the distributions requires also the knowledge
of the two-loop coefficient function, which at present is unknown for any
distribution.
After the resummation of the threshold logarithms in the $N$-space, 
we have returned to the form factor in the $x$-space,
by  inverting $f_N(\alpha_S)$ with  analytic and  numerical procedures.

\section{QCD form factor at NNLO}
\label{sec_computation}

Let us briefly describe the derivation of the NNLO form factor
in the $N$-space. By NNLO
accuracy, we mean the resummation of all the infrared logarithms up to and
including 
\begin{equation}
NNLO:\quad \alpha _{S}^{n}\,L^{n-1}.
\end{equation}
The general expression of the form factor is: 
\begin{equation}
\log f_{N}\left( \alpha _{S}\right) =\int_{0}^{1}dz\,\frac{z^{N-1}-1}{1-z}%
\left\{ \int_{Q^{2}(1-z)^{2}}^{Q^{2}(1-z)}\frac{dk^{2}}{k^{2}}\,A\left[
\alpha _{S}\left( k^{2}\right) \right] +D\left[ \alpha _{S}\left(
Q^{2}\left( 1-z\right) ^{2}\right) \right] +B\left[ \alpha _{S}\left(
Q^{2}(1-z)\right) \right] \right\} .  
\label{generale}
\end{equation}

Let us remind one difference between
annihilation processes, as Drell-Yan, and 
other processes, like the present one or DIS. 
In the annihilation processes,
 the initial partons reduce their momenta 
by irradiation, before actually annihilating;
 when $\tau = Q^2 /s \rightarrow 1$, only the emission 
of soft partons is allowed. Likewise, in the other processes, as f.i. DIS,
 and in the limit 
$x = Q^2 /2 Q\cdot p  \rightarrow 1$, only soft emission is allowed, before
the scattering; however, after the scattering, the parton 
fragments with the only 
kinematical constraint of having a low virtuality and collinear emission
is no longer forbidden. 

 The functions $A\left( \alpha _{S}\right) ,\,D\left( \alpha
_{S}\right) $ and $B\left( \alpha _{S}\right) $ have a perturbative
expansion in powers of $\alpha _{S}$:
\begin{eqnarray}
\qquad \qquad \qquad \qquad \qquad A\left( \alpha _{S}\right)
&=&\sum_{n=1}^{\infty }A_{n}\,\alpha _{S}^{n}=A_{1}\alpha _{S}+A_{2}\alpha
_{S}^{2}+A_{3}\alpha _{S}^{3}+\cdots ;  \notag \\
D\left( \alpha _{S}\right) &=&\sum_{n=1}^{\infty }D_{n}\,\alpha
_{S}^{n}=D_{1}\alpha _{S}+D_{2}\alpha _{S}^{2}+\cdots ;  \notag \\
B\left( \alpha _{S}\right) &=&\sum_{n=1}^{\infty }B_{n}\,\alpha
_{S}^{n}=B_{1}\alpha _{S}+B_{2}\alpha _{S}^{2}+\cdots .
\end{eqnarray}

Let us observe, that in general the transverse momentum rule is a guess, as
it has not been proven to such an accuracy. The universality of  $B_{2},$
in general, is a debated problem. If we neglect the variation of the
coupling with the scale (frozen coupling), we find logarithmic terms of the
form: 
\begin{equation}
A_{1}\alpha _{S}L^{2},\quad A_{2}\alpha _{S}^{2}L^{2},\quad A_{3}\alpha
_{S}^{3}L^{2},\cdots \quad D_{1}\alpha _{S}L,\quad D_{2}\alpha
_{S}^{2}L,\cdots \quad B_{1}\alpha _{S}L,\quad B_{2}\alpha _{S}^{2}L,\cdots .
\end{equation}
\newline
Then, to NNLO accuracy, one needs the first three terms in the expansion of $%
A\left( \alpha _{S}\right) $ and the first two terms of the functions $%
D\left( \alpha _{S}\right) $ and $B\left( \alpha _{S}\right) .$

The three-loop coupling, according to the definition given in the PDG \cite
{pdg}, reads: 
\begin{equation}
\alpha _{s}\left( \mu ^{2}\right) =\frac{1}{\beta _{0}\log \mu ^{2}/\Lambda
^{2}}-\frac{\beta _{1}}{\beta _{0}^{3}}\,\frac{\log \log \mu ^{2}/\Lambda
^{2}}{\log ^{2}\mu ^{2}/\Lambda ^{2}}+\frac{\beta _{1}^{2}}{\beta _{0}^{5}}\,%
\frac{\log ^{2}\log \mu ^{2}/\Lambda ^{2}-\log \log \mu ^{2}/\Lambda ^{2}-1}{%
\log ^{3}\mu ^{2}/\Lambda ^{2}}+\frac{\beta _{2}}{\beta _{0}^{4}}\frac{1}{%
\log ^{3}\mu ^{2}/\Lambda ^{2}}.  \label{coupling}
\end{equation}
The asymptotic expansion of the coupling is basically an expansion in
inverse powers of $\log \mu ^{2}/\Lambda ^{2}.$ The first three coefficients
of the $\beta $-function, defined as 
\begin{equation}
\frac{d\alpha _{s}}{d\log \mu ^{2}}=-\beta _{0}\alpha _{S}^{2}-\beta
_{1}\alpha _{S}^{3}-\beta _{2}\alpha _{S}^{4}-\cdots ,  \label{betafun}
\end{equation}
read: 
\begin{eqnarray}
\qquad \beta _{0} &=&\frac{11C_{A}-2n_{F}}{12\pi }=\frac{33-2n_{F}}{12\pi }%
=0.87535-0.05305\,n_{F},  \notag \\
\beta _{1} &=&\frac{17C_{A}^{2}-5C_{A}n_{F}-3C_{F}n_{F}}{24\pi ^{2}}=\frac{%
153-19\,n_{F}}{24\pi ^{2}}=0.64592-0.08021\,n_{F},  \notag \\
\beta _{2} &=&\frac{2857-5033/9\,n_{F}+325/27\,n_{F}^{2}}{128\pi ^{3}}%
=0.71986-0.140904\,n_{F}+0.003032\,n_{F}^{2}.
\end{eqnarray}
Let us note that $\beta _{0}$ and $\beta _{1}$ are renormalization-scheme
independent, while $\beta _{2}$ is not and we have given its value in the $%
\overline{MS}$ scheme \cite{beta2}.

Integrating
the $\beta $-function equation (\ref{betafun}) on both sides, one obtains: 
\begin{equation}
\alpha _{S}\left( k^{2}\right) =\alpha _{S}\left( Q^{2}\right) -\beta
_{0}\alpha _{S}^{2}\left( Q^{2}\right) \log \frac{k^{2}}{Q^{2}}-\beta
_{1}\alpha _{S}^{3}\left( Q^{2}\right) \log \frac{k^{2}}{Q^{2}}-\beta
_{2}\alpha _{S}^{4}\left( Q^{2}\right) \log \frac{k^{2}}{Q^{2}}+(\mathrm{%
iterations}).
\label{sviluppo}
\end{equation}
Substituting the above expression into (\ref{generale}), one sees that a $%
\beta _{0}$-insertion corresponds to the additional factor $\alpha _{S}\,L,$
that of $\beta _{1}$ to $\alpha _{S}^{2}\,L$ and that of $\beta _{2}$ to $%
\alpha _{S}^{3}\,L:$%
\begin{equation}
\beta _{0}:\alpha _{S}\,L,\qquad \beta _{1}:\alpha _{S}^{2}\,L,\qquad \beta
_{2}:\alpha _{S}^{3}\,L.
\end{equation}
Therefore, in the terms containing NNLO coefficients, the coupling can be
replaced with the one-loop one and in the NLO terms the coupling can be
replaced with the two-loop one, so that one has: 
\begin{eqnarray}
\qquad \qquad \qquad \qquad \qquad \qquad A\left( \alpha _{S}\right) 
&=&A_{1}\alpha _{S,3L}+A_{2}\alpha _{S,2L}^{2}+A_{3}\alpha _{S,1L}^{3}; 
\notag \\
D\left( \alpha _{S}\right)  &=&D_{1}\alpha _{S,2L}+D_{2}\alpha _{S,1L}^{2}; 
\notag \\
B\left( \alpha _{S}\right)  &=&B_{1}\alpha _{S,2L}+B_{2}\alpha _{S,1L}^{2}.
\label{furbata}
\end{eqnarray}
Furthermore, the $\beta _{1}^{2}$ term in $A_{2}\alpha _{S,2L}^{2}$ can be
neglected, as it is a N$^{3}$LO contribution. After replacing eqs.\thinspace
(\ref{furbata}) into eq.\thinspace (\ref{generale}), one performs a
straightforward integration over $k^{2},$ the gluon transverse momentum. The
integration over $z,$ the longitudinal gluon momentum, is easily done using
the approximation \cite{cattren}: 
\begin{equation}
z^{N-1}-1\simeq -\theta \left( 1-z-\frac{1}{n}\right) .  \label{ctapprox}
\end{equation}
This approximation misses the term proportional to $A_{1}\zeta _{2},$ where $%
\zeta _{2}=\pi ^{2}/6,$ which can be obtained in the following way. Using
the large-$N$ approximation derived in \cite{vogt}, one obtains, in the $n$
variable: 
\begin{eqnarray}
\int_{0}^{1}dz\,\frac{z^{N-1}-1}{1-z}\int_{Q^{2}(1-z)^{2}}^{Q^{2}(1-z)}\frac{%
dk^{2}}{k^{2}}\,A_{1}\alpha _{S}\left( k^{2}\right)  &=&\sum_{k=1}^{\infty
}\,c_{k}\int_{0}^{1}dz\,\frac{z^{N-1}-1}{1-z}\log ^{k}\left( 1-z\right)  
\notag \\
&\simeq &\sum_{k=1}^{\infty }c_{k}\,\frac{\left( -1\right) ^{k+1}}{k+1}\left[
\log ^{k+1}n+\frac{\zeta _{2}}{2}k\left( k+1\right) \log ^{k-1}n\right]  
\notag \\
&=&\left[ \mathrm{lo}\right] +\left[ \mathrm{nnlo}\right] .
\end{eqnarray}

Therefore: 
\begin{equation}
\left[ \mathrm{nnlo}\right] =\frac{\zeta _{2}}{2}\,\frac{\partial ^{2}}{%
\partial \left( \log n\right) ^{2}}\left[ \mathrm{lo}\right] .
\end{equation}

We have used the variable  $n=N/N_{0}$ instead of $%
N,$ where $N_{0}\equiv e^{-\gamma _{E}}=0.561459\ldots .$ Within this
alternative representation, the terms proportional to $\gamma _{E}$ and to $%
\gamma _{E}^{2}$ \ disappear. This scheme is probably more accurate as
Feynman diagram computation directly in $N$-space brings factors containing $%
\Gamma \left( 1-\epsilon \right) $ with $D=4-2\epsilon $ the space-time
dimension.

The advantage of  the variable $N$ is that the total rate is directly
reproduced by setting $N=1,$ while in the variable $n$ it is given by $%
f_{n=1/N_{0}}$. These two variables differ by terms of higher order in $%
\gamma _{E}.$

The lowest-order term, computed within the approximation (\ref{ctapprox}),
reads: 
\begin{equation}
\left[ \mathrm{lo}\right] =-\frac{A_{1}}{2\beta _{0}}\left[ \log \frac{s}{%
n^{2}}\log \log \frac{s}{n^{2}}+\log s\log \log s-2\log \frac{s}{n}\log \log 
\frac{s}{n}\right] .
\end{equation}
One then obtains: 
\begin{eqnarray}
\log f_{N}\left( \alpha _{S}\right)  &=&-\frac{A_{1}}{2\beta _{0}}\left[
\log s/n^{2}\log \log s/n^{2}+\log s\log \log s-2\log s/n\log \log s/n\right]
+  \notag \\
&&+\frac{\beta _{0}A_{2}-\beta _{1}A_{1}}{2\beta _{0}^{3}}\left[ \log \log
s-2\log \log s/n+\log \log s/n^{2}\right] +  \notag \\
&&-\frac{\beta _{1}A_{1}}{4\beta _{0}^{3}}\left[ \log ^{2}\log s/n^{2}-2\log
^{2}\log s/n+\log ^{2}\log s\right] +  \notag \\
&&+\frac{D_{1}}{2\beta _{0}}\left[ \log \log s/n^{2}-\log \log s\right] +%
\frac{B_{1}}{\beta _{0}}\left[ \log \log s/n-\log \log s\right]   \notag \\
&&-\frac{A_{3}}{4\beta _{0}^{3}}\left[ \frac{1}{\log s/n^{2}}-\frac{2}{\log
s/n}+\frac{1}{\log s}\right] -\frac{A_{1}\zeta _{2}}{2\beta _{0}}\left[ 
\frac{2}{\log s/n^{2}}-\frac{1}{\log s/n}-\frac{1}{\log s}\right] +  \notag
\\
&&-\frac{A_{1}\beta _{2}}{8\beta _{0}^{4}}\left[ \frac{1}{\log s/n^{2}}-%
\frac{2}{\log s/n}+\frac{1}{\log s}\right] +  \notag \\
&&+\frac{A_{2}\beta _{1}}{4\beta _{0}^{4}}\left[ 2\frac{\log \log s/n^{2}}{%
\log s/n^{2}}-4\frac{\log \log s/n}{\log s/n}+2\frac{\log \log s}{\log s}+%
\frac{3}{\log s/n^{2}}-\frac{6}{\log s/n}+\frac{3}{\log s}\right] +  \notag
\\
&&-\frac{A_{1}\beta _{1}^{2}}{4\beta _{0}^{5}}\left[ \frac{\log ^{2}\log
s/n^{2}}{\log s/n^{2}}-2\frac{\log ^{2}\log s/n}{\log s/n}+\frac{\log
^{2}\log s}{\log s}+2\frac{\log \log s/n^{2}}{\log s/n^{2}}+\right.   \notag
\\
&&\qquad \qquad \,\,\,\left. -4\frac{\log \log s/n}{\log s/n}+2\frac{\log
\log s}{\log s}+\frac{1}{\log s/n^{2}}-\frac{2}{\log s/n}+\frac{1}{\log s}%
\right] +  \notag \\
&&+\frac{D_{1}\beta _{1}}{2\beta _{0}^{3}}\left[ \frac{\log \log s/n^{2}}{%
\log s/n^{2}}-\frac{\log \log s}{\log s}+\frac{1}{\log s/n^{2}}-\frac{1}{%
\log s}\right] -\frac{D_{2}}{2\beta _{0}^{2}}\left[ \frac{1}{\log s/n^{2}}-%
\frac{1}{\log s}\right] +  \notag \\
&&+\frac{B_{1}\beta _{1}}{\beta _{0}^{3}}\left[ \frac{\log \log s/n}{\log s/n%
}-\frac{\log \log s}{\log s}+\frac{1}{\log s/n}-\frac{1}{\log s}\right] -%
\frac{B_{2}}{\beta _{0}^{2}}\left[ \frac{1}{\log s/n}-\frac{1}{\log s}\right]
.
\end{eqnarray}
The next step is to write the above result as a function of the three-loop
coupling. The inverse of (\ref{coupling})\ reads: 
\begin{equation}
\log s=\frac{1}{\beta _{0}\alpha _{S}}+\frac{\beta _{1}}{\beta _{0}^{2}}\log
\left( \beta _{0}\alpha _{S}\right) -\left( \frac{\beta _{1}^{2}}{\beta
_{0}^{3}}-\frac{\beta _{2}}{\beta _{0}^{2}}\right) \alpha _{S}+O\left(
\alpha _{S}^{2}\right) ,  \label{logsrep}
\end{equation}
where $s$ is the square of the hard scale in unit of the QCD scale, 
\begin{equation}
s\equiv \frac{Q^{2}}{\Lambda ^{2}}.
\end{equation}
After the replacement (\ref{logsrep}), terms of higher order with respect to
NNLO are generated, which must be discarded. One finally makes the expansion
in $\gamma _{E}$ up to second order to obtain eq.\thinspace (\ref{saragiusta}%
).

The quantities $A_{1}$ and $A_{2}$ are known analytically \cite
{cattren,kodtren}: 
\begin{equation}
A_{1}=\frac{C_{F}}{\pi }=0.424413,\qquad A_{2}=\frac{C_{F}}{\pi ^{2}}\left[
C_{A}\left( \frac{67}{36}-\frac{\pi ^{2}}{12}\right) -\frac{5}{9}n_{F}T_{R}%
\right] =0.42095-0.03753\,n_{F},
\end{equation}
where $C_{A}=N_{c}=3,$ $T_{R}=1/2$ and $n_{F}=3$ is the number of active
quark flavours in $b$ decay. The value given for $A_{2}$ in the $\overline{MS%
}$ \ scheme for the coupling constant. At present, only a numerical estimate
of $A_{3}$ is available \cite{vogt, vaneerven}: 
\begin{equation}
A_{3}=0.59413-0.09272\,n_{F}-0.00040\,n_{F}^{2}.
\end{equation}
The soft quantities $D_{1}$ and $D_{2}$ \cite{kor} are known analitically: 
\begin{eqnarray}
D_{1} &=&-\frac{C_{F}}{\pi }=-0.424413, \\
D_{2} &=&-\frac{C_{F}}{\pi ^{2}}\left[ \left( \frac{37}{108}+\frac{7}{18}\pi
^{2}-\frac{9}{4}\zeta \left( 3\right) \right) C_{A}+\left( \frac{1}{27}-%
\frac{1}{9}\pi ^{2}\right) T_{R} \, n_{F}\right] =-0.59826-0.07157%
\,n_{F}.  \notag
\end{eqnarray}
Numerically, $\zeta \left( 3\right) \cong 1.20206.$ The constant $D_{1}$ is
renormalization-scheme independent, while $D_{2}$ is not and we have given
its value in the $\overline{MS}$  scheme. The latter quantity turns out to
be the most important one to determine the NNLO effects.

The collinear quantity $B_{1}$ is known analitically, 
\begin{equation}
B_{1}=-\frac{3}{4}\frac{C_{F}}{\pi }=-0.31831.
\end{equation}
The two-loop quantity $B_{2}$ is unknown and we approximate it with the
infrared-regular part in the two-loop Altarelli-Parisi splitting function $%
P_{qq}\left( z\right)$. In general, the latter is naturally decomposed in
the soft limit as: 
\begin{equation}
P_{qq}\left( z\right) =\left[ \frac{A(\alpha _{S}\left( Q^{2}(1-z)\right) )}{%
1-z}\right] _{+}-K\left( z;\alpha _{S}\right) +K\left( \alpha _{S}\right)
\,\delta \left( 1-z\right) .
\end{equation}
The functions $K\left( z;\alpha _{S}\right) $ and $K\left( \alpha
_{S}\right) $ have an expansion in powers of $\alpha _{S}$: 
\begin{eqnarray}
K\left( \alpha _{S}\right)  &=&K_{1}\alpha _{S}+K_{2}\alpha _{S}^{2}+\cdots ,
\notag \\
K\left( z;\alpha _{S}\right)  &=&K_{1}\left( z\right) \alpha
_{S}+K_{2}\left( z\right) \alpha _{S}^{2}+\cdots .
\end{eqnarray}
Since the splitting function has vanishing first moment: 
\begin{equation}
K\left( \alpha _{S}\right) =\int_{0}^{1}K\left( z;\alpha _{S}\right) dz,
\end{equation}
it holds that 
\begin{equation}
K_{1}=\int_{0}^{1}K_{1}\left( z\right) dz\quad \quad \mathrm{and}\qquad
K_{2}=\int_{0}^{1}K_{2}\left( z\right) dz.
\end{equation}
In leading order, it holds $B_{1}=K_{1}.$ Our approximation then is:\
\bigskip\ 
\begin{equation}
B_{2}\approx K_{2}  \label{rozza}
\end{equation}
where, in the $\overline{MS}$ scheme
\cite{kor2}, 
\begin{eqnarray}
K_{2} &=&-\frac{C_{F}}{\pi ^{2}}\left[ C_{F}\left( \frac{3}{32}-\frac{\pi
^{2}}{8}+\frac{3}{2}\zeta \left( 3\right) \right) +C_{A}\left( \frac{17}{%
96}+\frac{11}{72}\pi ^{2}-\frac{3}{4}\zeta \left( 3\right) \right)
-\left( \frac{1}{24}+\frac{\pi ^{2}}{18}\right) T_{R} \, n_{F}\right]
   \notag \\
&=&-0.43695+0.03985\,n_{F}.
\end{eqnarray}

\subsection{Results}

The functions $%
g_{1}$ and $g_{2}$ have the following expressions \cite{americani,subleading}: 
\begin{eqnarray}
g_{1}\left( \lambda ;\frac{\mu ^{2}}{Q^{2}}\right)  &=&-\frac{A_{1}}{2\beta
_{0}}\,\frac{1}{\lambda }\left[ \left( 1-2\lambda \right) \log \left(
1-2\lambda \right) -2\left( 1-\lambda \right) \log \left( 1-\lambda \right) %
\right] ;  \label{compatta1} \\
\,g_{2}\left( \lambda ;\frac{\mu ^{2}}{Q^{2}}\right)  &=&+\frac{A_{2}}{%
2\beta _{0}^{2}}\left[ \log (1-2\lambda )-2\log (1-\lambda )\right] +\frac{%
A_{1}\gamma _{E}}{\beta _{0}}\left[ \log (1-2\lambda )-\log (1-\lambda )%
\right] +  \notag \\
&&-\frac{\beta _{1}A_{1}}{4\beta _{0}^{3}}\left[ \log ^{2}(1-2\lambda
)-2\log ^{2}(1-\lambda )+2\log (1-2\lambda )-4\log (1-\lambda )\right] + 
\notag \\
&&+\frac{D_{1}}{2\beta _{0}}\log (1-2\lambda )+\frac{B_{1}}{\beta _{0}}\log
(1-\lambda )+\frac{A_{1}}{2\beta _{0}}\,\left[ \log \left( 1-2\lambda
\right) -2\log \left( 1-\lambda \right) \right] \log \frac{\mu ^{2}}{Q^{2}}.
\label{compatta2}
\end{eqnarray}
Our result for the NNLO function $\,g_{3}$\ reads: 
\begin{eqnarray}
g_{3}\left( \lambda ;\frac{\mu ^{2}}{Q^{2}}\right)  &=&-\frac{A_{3}}{2\beta
_{0}^{2}}\left[ \frac{\lambda }{1-2\lambda }-\frac{\lambda }{1-\lambda }%
\right] -\frac{A_{1}\zeta _{2}}{2}\left[ \frac{4\lambda }{1-2\lambda }-\frac{%
\lambda }{1-\lambda }\right] +  \notag \\
&&-\frac{A_{1}\beta _{2}}{4\beta _{0}^{3}}\left[ \frac{2\lambda }{1-2\lambda }%
-\frac{2\lambda }{1-\lambda }+\log \left( 1-2\lambda \right) -2\log \left(
1-\lambda \right) \right] +  \notag \\
&&+\frac{A_{2}\beta _{1}}{2\beta _{0}^{3}}\left[ \frac{\log \left(
1-2\lambda \right) }{1-2\lambda }-\frac{2\log \left( 1-\lambda \right) }{%
1-\lambda }+\frac{3\lambda }{1-2\lambda }-\frac{3\lambda }{1-\lambda }\right]
+  \notag \\
&&-\frac{A_{1}\beta _{1}^{2}}{2\beta _{0}^{4}}\left[ \frac{1}{2}\frac{\log
^{2}\left( 1-2\lambda \right) }{1-2\lambda }-\frac{\log ^{2}\left( 1-\lambda
\right) }{1-\lambda }+\frac{\log \left( 1-2\lambda \right) }{1-2\lambda }%
+\right.   \notag \\
&&\left. -\frac{2\log \left( 1-\lambda \right) }{1-\lambda }+\frac{\lambda }{%
1-2\lambda }-\frac{\lambda }{1-\lambda }-\log \left( 1-2\lambda \right)
+2\log \left( 1-\lambda \right) \right] +  \notag \\
&&+\frac{D_{1}\beta _{1}}{2\beta _{0}^{2}}\left[ \frac{\log \left(
1-2\lambda \right) }{1-2\lambda }+\frac{2 \lambda }{1-2\lambda }\right] +%
\frac{B_{1}\beta _{1}}{\beta _{0}^{2}}\left[ \frac{\log \left( 1-\lambda
\right) }{1-\lambda }+\frac{\lambda }{1-\lambda }\right]+   \notag \\
&&-\frac{D_{2}}{\beta _{0}}\frac{\lambda }{1-2\lambda }-\frac{B_{2}}{\beta
_{0}}\frac{\lambda }{1-\lambda }-\frac{A_{1}\gamma _{E}^{2}}{2}
\left[ \frac{4 \lambda }{1-2\lambda }-
\frac{\lambda}{1-\lambda }\right] +  \notag \\
&&+\frac{A_{1}\beta _{1}\gamma _{E}}{\beta _{0}^{2}}\left[ \frac{\log \left(
1-2\lambda \right) }{1-2\lambda }-\frac{\log \left( 1-\lambda \right) }{%
1-\lambda }+\frac{1}{1-2\lambda }-\frac{1}{1-\lambda }\right] +  \notag \\
&&-\frac{A_{2}\gamma _{E}}{\beta _{0}}\left[ \frac{1}{1-2\lambda }-\frac{1}{%
1-\lambda }\right] -\frac{D_{1}\gamma _{E} 2 \lambda}{1-2\lambda }-\frac{B_{1}\gamma
_{E} \lambda}{1-\lambda }+  \notag \\
&&-\frac{A_{1}}{2\beta _{0}}\left[ \frac{2\lambda ^{2}}{1-2\lambda }-\frac{%
\lambda ^{2}}{1-\lambda }\right] \log ^{2}\frac{\mu ^{2}}{Q^{2}}-\frac{A_{2}%
}{\beta _{0}^2}\left[ \frac{\lambda }{1-2\lambda }-\frac{\lambda }{1-\lambda }%
\right] \log \frac{\mu ^{2}}{Q^{2}}+  \notag \\
&&-\frac{A_{1}\gamma _{E}}{\beta _{0}}\left[ \frac{2\lambda }{1-2\lambda }-%
\frac{\lambda }{1-\lambda }\right] \log \frac{\mu ^{2}}{Q^{2}}-\frac{D_{1}}{%
\beta _{0}}\frac{\lambda }{1-2\lambda }\log \frac{\mu ^{2}}{Q^{2}}-\frac{%
B_{1}}{\beta _{0}}\frac{\lambda }{1-\lambda }\log \frac{\mu ^{2}}{Q^{2}}+ 
\notag \\
&&+\frac{A_{1}\beta _{1}}{\beta _{0}^{3}}\left[ \frac{\lambda \log \left(
1-2\lambda \right) }{1-2\lambda }-\frac{\lambda \log \left( 1-\lambda
\right) }{1-\lambda }+\frac{\lambda }{1-2\lambda } + \right. \notag \\
&& \left. -\frac{\lambda }{1-\lambda 
}  + \frac{1}{2} \log (1-2 \lambda) - \log (1-\lambda)\right] 
\log \frac{\mu ^{2}}{Q^{2}}.  \label{saragiusta}
\end{eqnarray}
Arbitrary constants have been added to the function
$g_3$ in order to make it homogenous.
The quantity $\gamma _{E}=0.577216\ldots $ is the Euler constant and $\zeta
\left( n\right) $ is the Rieman zeta function, 
\begin{equation}
\zeta \left( n\right) \equiv \sum_{k=1}^{\infty }\frac{1}{k^{n}}.
\end{equation}
$\zeta \left( 2\right) =\pi ^{2}/6=1.64493.$ The functions $g_{2}$ and $g_{3}
$ depend on the renormalization scale $\mu ,$ while $g_{1}$ does not.

The expansion
up to order $\alpha _{S}^{3}$ reads:
\begin{eqnarray}
\log f_{N} &=&-\frac{1}{2}A_{1}\alpha _{S}L^{2}-D_{1}\alpha
_{S}L-B_{1}\alpha _{S}L+  \notag \\
&&-\frac{1}{2}A_{1}\beta _{0}\,\alpha _{S}^{2}\,L^{3}-\left( \frac{1}{2}%
A_{2}+\beta _{0}D_{1}+\frac{1}{2}\beta _{0}B_{1}\right) \alpha
_{S}^{2}\,L^{2}-\left( D_{2}+B_{2}+\frac{3}{2}A_{1}\zeta _{2}\beta
_{0}\right) \alpha _{S}^{2}L+  \notag \\
&&-\frac{7}{12}A_{1}\beta _{0}^{2}\alpha _{S}^{3}L^{4}-\left( A_{2}\beta
_{0}+\frac{1}{2}A_{1}\beta _{1}+\frac{4}{3}D_{1}\beta _{0}^{2}+\frac{1}{3}%
B_{1}\beta _{0}^{2}\right) \alpha _{S}^{3}L^{3}+  \notag \\
&&-\left( \frac{1}{2}A_{3}+\frac{7}{2}A_{1}\zeta _{2}\beta
_{0}^{2}+2D_{2}\beta _{0}+B_{2}\beta _{0}+D_{1}\beta _{1}+\frac{1}{2}%
B_{1}\beta _{1}\right) \alpha _{S}^{3}L^{2}+O\left( \alpha _{S}^{4}\right) .
\label{NNLOfN}
\end{eqnarray}
Let us note that $\beta _{2}$ appears only at order $O\left( \alpha
_{S}^{4}\right) .$

The functions $g_{i}$ become singular when 
\begin{equation}
\lambda \rightarrow \frac{1}{2}^{-}.
\label{firstLP}
\end{equation}
Since 
\begin{equation}
\lambda =\beta _{0}\alpha _{S}\left( \mu ^{2}\right) L,
\end{equation}
this means that a singularity in $N$-space occurs when: 
\begin{equation}
N\rightarrow \exp \left[ \frac{1}{2\beta _{0}\alpha _{S}\left( \mu
^{2}\right) }\right] \approx \frac{\mu }{\Lambda },
\end{equation}
where $\Lambda $ is the QCD scale. Let us observe that, in general, this
singularity signals non-perturbative effects but its precise position is
completely unphysical, as we can move it with a change of renormalization
scale.

\subsection{Renormalization-scale dependence}
\label{ren-scale}

In this section we consider renormalization-scale dependence. In principle,
such scale $\mu $ should not appear in the cross sections, as it does not
correspond to any fundamental constant or kinematical scale in the problem.
The completely resummed perturbative expansion of an observable\footnote{%
We do not mean here the resummation of towers of logarithmic contributions,
but the resummation of the whole series in $\alpha _{S}.$} is indeed
formally independent on $\mu .$ In practise, truncated perturbative
expansions exhibit a residual scale dependence, because of neglected higher
orders.

We start with the form factor as a function of $\alpha _{S}\left(
Q^{2}\right) $ and we derive its expression as a function of $\alpha
_{S}\left( \mu ^{2}\right) $ and $\mu ^{2}/Q^{2}.$

\begin{figure}[b]
\par
\begin{center}
\epsfxsize=0.80\textwidth
\epsfysize=0.50\textheight
\leavevmode\epsffile{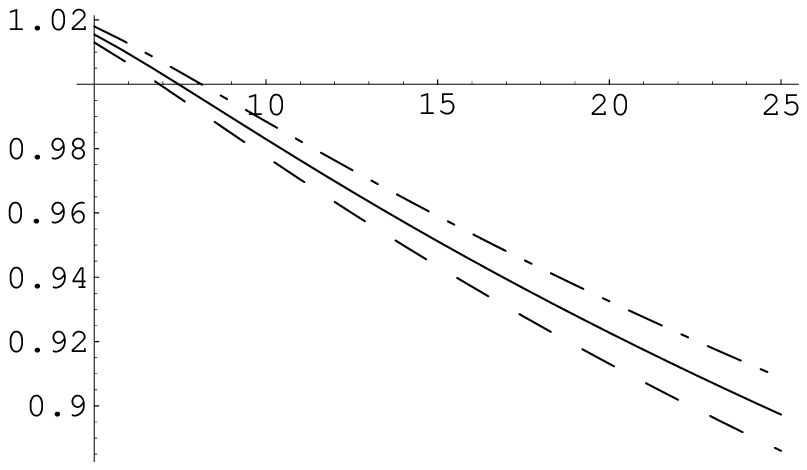}
\end{center}
\caption{{\rm{ We show the dependence of $\log[N]$
on the scale $\mu$ at NLO, with $\mu \simeq m_B$ (solid line),
$\mu \simeq 2 m_B$ (dashed line) and
$\mu \simeq m_B/2$ (dot-dashed  line),
 at fixed $\alpha_S(Q^2)=0.21$  and $5<N<25$.}}}
\label{fig_scaling1}
\end{figure}

\begin{figure}[b]
\par
\begin{center}
\epsfxsize=0.80\textwidth
\epsfysize=0.50\textheight
\leavevmode\epsffile{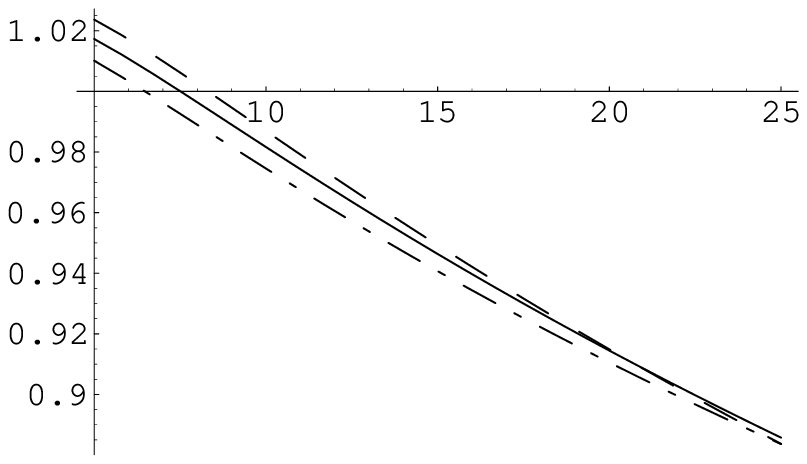}
\end{center}
\caption{{\rm{ We show the dependence of $\log[N]$
on the scale $\mu$ at NNLO, with $\mu \simeq m_B$ (solid line),
$\mu \simeq 2 m_B$ (dashed line) and
$\mu \simeq m_B/2$ (dot-dashed  line),
 at fixed $\alpha_S(Q^2)=0.21$  and $5<N<25$.}}}
\label{fig_scaling2}
\end{figure}

Since 
\begin{equation}
a\left( Q\right) =a\left( \mu \right) +c\,a^{2}\left( \mu \right) +c^{\prime
}\,a^{3}\left( \mu \right) +O\left( a^{4}\right) ,
\end{equation}
where $a\equiv \beta _{0}\alpha _{S}$ and (from eq. (\ref{sviluppo}))
\begin{equation}
c=\log \frac{\mu ^{2}}{Q^{2}},\qquad \quad c^{\prime }=\log ^{2}\frac{\mu
^{2}}{Q^{2}}+\frac{\beta _{1}}{\beta _{0}^{2}}\log \frac{\mu ^{2}}{Q^{2}},
\end{equation}
one has, where now $\lambda = \lambda(\mu)$: 
\begin{eqnarray}
L\,g_{1}\left[ \lambda +ac\lambda +a^{2}c^{\prime }\lambda \right] 
&=&L\,g_{1}\left[ \lambda \right] +c\lambda ^{2}g_{1}^{\prime }\left[
\lambda \right] +ac^{\prime }\,\lambda ^{2}g_{1}^{\prime }\left[ \lambda %
\right] +a\frac{1}{2}c^{2}\,\lambda ^{3}g_{1}^{\prime \prime }\left[ \lambda %
\right] +\cdots   \notag \\
g_{2}\left[ \lambda +ac\lambda +a^{2}c^{\prime }\lambda \right]  &=&\,g_{2}%
\left[ \lambda \right] +a\,c\lambda \,g_{2}^{\prime }\left[ \lambda \right]
+\cdots .
\end{eqnarray}
The additional terms in the functions $g_{i},$ to (partially) compensate for
the scale change $Q^{2}\rightarrow \mu ^{2}$ therefore read: 
\begin{eqnarray}
\delta g_{1}\left[ \lambda ,\frac{\mu ^{2}}{Q^{2}}\right]  &=&0  \notag \\
\delta g_{2}\left[ \lambda ,\frac{\mu ^{2}}{Q^{2}}\right]  &=&\lambda
^{2}g_{1}^{\prime }\left[ \lambda \right] \log \frac{\mu ^{2}}{Q^{2}}  \notag
\\
\delta g_{3}\left[ \lambda ,\frac{\mu ^{2}}{Q^{2}}\right]  &=&\frac{1}{2}%
\lambda ^{3}g_{1}^{\prime \prime }\left[ \lambda \right] \log ^{2}\frac{\mu
^{2}}{Q^{2}}+\lambda ^{2}g_{1}^{\prime }\left[ \lambda \right] \left( \log
^{2}\frac{\mu ^{2}}{Q^{2}}+\frac{\beta _{1}}{\beta _{0}^{2}}\log \frac{\mu
^{2}}{Q^{2}}\right) +\lambda \,g_{2}^{\prime }\left[ \lambda \right] \log 
\frac{\mu ^{2}}{Q^{2}}  \notag \\
&=&\frac{1}{2}\lambda \frac{d}{d\lambda }\left( \lambda ^{2}\,g_{1}^{\prime }%
\left[ \lambda \right] \right) \,\log ^{2}\frac{\mu ^{2}}{Q^{2}}+\left( 
\frac{\beta _{1}}{\beta _{0}^{2}}\lambda ^{2}g_{1}^{\prime }\left[ \lambda %
\right] +\lambda \,g_{2}^{\prime }\left[ \lambda \right] \right) \log \frac{%
\mu ^{2}}{Q^{2}}.
\label{gi_scaling}
\end{eqnarray}
Therefore, the leading function $g_{1}$ does not 
depend explicitely on the renormalization scale,
analogously to the coefficient of the leading term in the expansion of
inclusive observables (such as the $\alpha _{S}/\pi $ term in the total $%
e^{+}e^{-}$ hadronic cross-section). The higher order functions $g_{i>1}$
instead explicitly depend on $\mu $. We want to show the effect 
on $\log[N]$ of the
variations in eq. (\ref{gi_scaling}). Therefore,
in fig. \ref{fig_scaling1} and in fig.
\ref{fig_scaling2}, we show the
 dependence of $\log[N]$ on $\mu$ at NLO and at NNLO, 
keeping $\alpha_S$ fixed at 0.21; we notice a NNLO 
improvement at high values of N.
At lower values of $N$ ($N\le 10$), there is an appreciable improvement
only by  going at  much higher energies (f.i. $\alpha_S = 0.1$). 
In general, by varying also $\alpha_S$ with the scale $\mu$ around $m_b$,
 there is not a significant improvement at NNLO. 

\section{Analytic inverse Mellin transform}
\label{SecInv}

The $N$-moments are physical quantities, but in practice a measure of the
moments for large $N$ is difficult. It is therefore convenient to perform the
inverse transform back to momentum space.
 

Let us derive an analytical expression of the inverse Mellin transform at
NNLO.
We start by introducing  the  inverse Mellin transform of 
$f_N (\alpha_S) / N$:
\begin{equation}
G (\alpha_S;x) = \frac{1}{2 \pi i }
\int_{C-i \infty}^{C+i \infty}\frac{dN}{N}\,x^{-N} f_N(\alpha_S, L)
\end{equation}

The inverse Mellin transform of $f_N (\alpha_S)$
is the logarithmic derivative of $G(\alpha_S, L;x)$:
\beq
f(x)=-x \,\frac{d}{dx} G(x).
\end{equation}

After the substitutions
$x \equiv e^{-s}$ and $u \equiv N s$,
we have:
\begin{eqnarray}
G (\alpha_S;x) &=& \frac{1}{2 \pi i }
\int_{C^\prime-i \infty}^{C^\prime+i \infty}\frac{du}{u}
\,e^{u} \, f_N(\alpha_S, L) \notag \\
&=&
 \frac{1}{2 \pi i }
\int_{C^\prime-i \infty}^{C^\prime+i \infty}\frac{du}{u}
\,e^{u} \,
\exp \left[ L\,g_{1}\left( \beta
_{0}\alpha _{S}L\right) +\sum_{n=2}^{\infty }\alpha _{S}^{n-2}\,g_{n}\left(
\beta _{0}\alpha _{S}L\right) \right]
\end{eqnarray}

We Taylor expand the exponent with respect to $L$ around $L=l=\ln 1/s$
\cite{thrust}. In the $x$-variable, 
$l\equiv-\ln(-\ln x)$. Note that 
$l \rightarrow -\ln(1-x)$ when $x \rightarrow 1$.

We have, at NNLO:
\beq
G(\alpha_S;x)= \frac{1}{2 \pi \, i }
\int_{C^\prime-i \infty}^{C^\prime+i \infty}\,
\frac{du}{u}\,e^{u+ F_0(l) + F_1(l) \ln u +\frac{1}{2}\, F_2(l)\, \ln^2 u +
...}
\end{equation}
where, at scale $Q^2$ and at the same NNLO:
\begin{eqnarray}
F_0(l) &=& l\, g_1(\beta_0 \alpha_S l) + g_2( \beta_0 \alpha_S l) +
\alpha_S \, g_3 (\beta_0 \alpha_S l), \nonumber \\
F_1(l) &=& g_1(\beta_0 \alpha_S l) + \beta_0 \alpha_S \, l 
\, g_1^\prime(\beta_0 \alpha_S l) +
 \beta_0 \alpha_S  \, g_2^\prime(\beta_0 \alpha_S l),  \nonumber \\
F_2(l) &=& 2 \beta_0 \alpha_S 
\, g_1^\prime(\beta_0 \alpha_S l) +
\beta_0^2 \alpha_S^2\, l  \, g_1^{\prime\, \prime}(\beta_0 \alpha_S l).
  \nonumber
\end{eqnarray}

By keeping in the exponent terms up to NLO, while expanding
the NNLO terms, we obtain:
\beq
G(\alpha_S;x)= \frac{ e^{F_0(l)}}{2 \pi  i }\,
\int_{C^\prime-i \infty}^{C^\prime+i \infty}\,
du\,e^{u -\left[1-F_1^{NL}(l)\right] \ln u }
\left[1+F_1^{N^2 L}(l)\, \ln u + \frac{1}{2}\, F_2(l)\, \ln^2 u 
+ \cdots \right]
\label{expansion_an}
\end{equation}
where:
\begin{eqnarray}
F_1(l) &\equiv& F_1^{NL}(l) + F_1^{N^2 L}(l) \notag \\
 F_1^{NL}(l) &\equiv&
 g_1(\beta_0 \alpha_S l) + \beta_0 \alpha_S \, l 
\, g_1^\prime(\beta_0 \alpha_S l) \notag \\ 
F_1^{N^2 L}(l) &\equiv& 
 \beta_0 \alpha_S  \, g_2^\prime(\beta_0 \alpha_S l).
\end{eqnarray}
By using the result:
\beq
\frac{1}{2 \pi \, i }
\int_{C}\,
du\,\log^k u \,e^{u-\left[1-F_1^{NL}(l)\right]\log u}
=\frac{d^k}{{d F_1^{NL}}^k}\, \frac{1}{\Gamma(1-F_1^{NL})},
\end{equation}
where $\Gamma$ is the Euler Gamma function, we obtain,
after the integration:
\begin{eqnarray}
G(\alpha_S;x) &=&   \frac{e^{F_0(l)}}{\Gamma(1-F_1^{NL})}\left[
1 + {F_1^{N^2 L}}\, \psi (1- {F_1^{NL}}) + \frac{1}{2}\, F_2(l)\,
\left(\psi^2(1- {F_1^{NL}}) - \psi^\prime(1- {F_1^{NL}})
\right)\right] \notag \\
&=& 
\frac{e^{l\, g_1(\beta_0 \alpha_S l) + g_2( \beta_0 \alpha_S l) +
\alpha_S \, g_3 (\beta_0 \alpha_S l)}}{
\Gamma( 1- g_1(\beta_0 \alpha_S l) - \beta_0 \alpha_S \, l 
\, g_1^\prime(\beta_0 \alpha_S l))} 
\left[1+ 
 \beta_0 \alpha_S  \, g_2^\prime(\beta_0 \alpha_S l) \,
 \psi ( 1- g_1(\beta_0 \alpha_S l) - \beta_0 \alpha_S \, l 
\, g_1^\prime(\beta_0 \alpha_S l)) \right. \notag \\ 
& & \left.
+\frac{1}{2}\, F_2(l)\,
\left(\psi^2 ( 1- g_1(\beta_0 \alpha_S l) - \beta_0 \alpha_S \, l 
\, g_1^\prime(\beta_0 \alpha_S l))-
 \psi^\prime( 1- g_1(\beta_0 \alpha_S l) - \beta_0 \alpha_S \, l 
\, g_1^\prime(\beta_0 \alpha_S l))
\right)\right] 
\end{eqnarray}
with $\psi(x)= d\log\Gamma(x)/dx$, the digamma function.

At the end, we find the explicit analytic formula for the inverse
 Mellin transform at NNLO:
\begin{eqnarray}
f(x) &=& -x \frac{d}{dx}\, G(\alpha_S;x) \nonumber \\
&=& -x \frac{d}{dx}\,\left\{
\frac{e^{l\, g_1(\beta_0 \alpha_S l) + g_2( \beta_0 \alpha_S l) +
\alpha_S \, g_3 (\beta_0 \alpha_S l)}}{
\Gamma( 1- g_1(\beta_0 \alpha_S l) - \beta_0 \alpha_S \, l 
\, g_1^\prime(\beta_0 \alpha_S l))} 
\left[1+ 
 \beta_0 \alpha_S  \, g_2^\prime(\beta_0 \alpha_S l) \,
 \psi ( 1- g_1(\beta_0 \alpha_S l) - \beta_0 \alpha_S \, l 
\, g_1^\prime(\beta_0 \alpha_S l)) \right. \right.\notag \\ 
& & \left. \left.
+\frac{1}{2}\, F_2(l)\,
\left(\psi^2 ( 1- g_1(\beta_0 \alpha_S l) - \beta_0 \alpha_S \, l 
\, g_1^\prime(\beta_0 \alpha_S l))-
 \psi^\prime( 1- g_1(\beta_0 \alpha_S l) - \beta_0 \alpha_S \, l 
\, g_1^\prime(\beta_0 \alpha_S l))
\right)\right]\right\}. 
\label{analytic_formula_finale}
\end{eqnarray}

\subsection{Discussion and conclusions}
\label{frozen}

There are two physical effects involved in the inverse 
Mellin transform:

\begin{enumerate}

\item  \textit{exact longitudinal momentum conservation;} the terms which
enforce longitudinal momentum conservation are formally subleading in the
infrared logarithm counting;

\item  \textit{infrared pole in the coupling;} the inverse transform
involves an integration over all moments and arbitrarily large values of $|N|
$ enter. This implies that a prescription for the infrared pole in the
coupling has to be given. In general, the transformation mixes all the
momentum scales in the problem.
\end{enumerate}

One can study one problem at a time; for example, one can study the problem
1. only, by considering the frozen coupling approximation for the
distributions in $N$-space.
 In one-loop, this is equivalent to consider the
QED case. 

The frozen coupling approximation means neglecting the variation of
 $\alpha_S$ with the scale.
Let us start from formula (\ref{generale}) at NNLO: 
\begin{eqnarray}
\log f_{N}\left( \alpha _{S}\right) &=&
\int_{0}^{1}dz\,\frac{z^{N-1}-1}{1-z}
\left\{ 
 \int_{Q^{2}(1-z)^{2}}^{Q^{2}(1-z)}\frac{dk^{2}}{k^{2}} \,\left[
A_1 \alpha_{S}+ A_2 \alpha_{S}^2 + A_3 \alpha_{S}^3 + ... \right] + \right.
\nonumber \\
 &+& \left.
B_1 \alpha_{S} + B_2 \alpha_{S}^2 + ...  +
D_1 \alpha_{S} + D_2 \alpha_{S}^2 + ...  +  \right\}   
\nonumber \\
&\simeq&
\int_{0}^{1} dz \frac{z^{N-1}-1}{1-z}
\left\{ \left( A_1 \alpha_{S}+ A_2 \alpha_{S}^2 + A_3 \alpha_{S}^3 \right)
\ln \frac{1}{1-z} + \left( B_1  + D_1 \right) \alpha_{S} +
\left( B_2  + D_2 \right) \alpha_{S}^2  \right\} \nonumber \\
& & 
\end{eqnarray}
After integration in $z$, at the lowest order in 
$\lambda= \beta_0 \alpha_2 \ln N$, we have:
\begin{eqnarray}
g_1 &=& - \frac{A_1}{2 \, \beta_0} \, \lambda\\
g_2 &=& 
\left( -\frac{B_1}{\beta_0} -\frac{D_1}{\beta_0}
-\frac{A_1 \gamma_E}{\beta_0} \right) \, \lambda\\
g_3 &=& \left( -\frac{B_2}{\beta_0} -\frac{D_2}{\beta_0}
-\frac{A_2 \gamma_E}{\beta_0} \right) \, \lambda.
\end{eqnarray}

We have two ways, analytical and numerical, to 
compute the  inverse Mellin transform of 
$f_N(\alpha_S)$:
\begin{equation}
f(\alpha_S;x) = \frac{1}{2 \pi i }
\int_{C-i \infty}^{C+i \infty} dN \,x^{-N} f_N(\alpha_S, L).
\label{MellinInverse} 
\end{equation}
In the case of frozen coupling, the $g_i$ are linear in $\lambda$
and therefore
the numerical path does not include the Landau pole;
the numerical integration
 is therefore  exact.
We have compared the inverse
Mellin transform 
calculated numerically, directly from the 
definition (\ref{MellinInverse}),
with the analytical result (\ref{analytic_formula_finale}).
We find a very good agreement, up to energies around the charm scale, provided
that we go to NNLO.
The exact longitudinal momentum conservation
does not seem to spoil the reliability of the perturbative series.

In the real world, we cannot use 
linearized $g_i$, but we have to use the $g_i$ computed in section 
\ref{sec_computation}; therefore, 
problems due to infrared poles come into play.

Let us first make an important observation: the degree of singularity of the
functions $g_{i}$ for $\lambda \rightarrow 1/2,$ and therefore also of the
form factor, increases with the order of the function, i.e. with $i$ \cite
{gardi}.

At LO, $g_{1}$ has at most a logarithmic singularity of the form 
\begin{equation}
g_{1}:\quad \left( 1-2\lambda \right) \log \left( 1-2\lambda \right) ;
\end{equation}
at NLO, $g_{2}$ has at most a singularity of the form 
\begin{equation}
g_{2}:\quad \log ^{2}\left( 1-2\lambda \right) .
\end{equation}
The prefactor $1-2\lambda $ in front of the logarithm, vanishing for $%
\lambda =1/2,$ is now absent, but the singularity is still logarithmic.
At
NNLO, $g_{3}$ has at most a singularity of the form 
\begin{equation}
g_{3}:\quad \frac{\log ^{2}\left( 1-2\lambda \right) }{1-2\lambda }.
\end{equation}
The latter is basically a pole singularity, no  more a logarithmic
singularity anymore, i.e. a much stronger singularity. Note that the above
term is proportional to $A_{1}\beta _{1}^{2}$ and so it is
scheme-independent.

Another observation concerns the size of the two-loop soft term. For three
active flavours, the size of the two-loop correction with respect to the
one-loop one is rather large, 
\begin{equation}
\frac{D_{2}}{D_{1}}\alpha _{S}\,\approx \,2\,\alpha _{S}\,\approx \,40\%,
\end{equation}
since $\alpha _{S}\left( m_{B}\right) \simeq 0.21.$ The inclusion of $D_{2}$
is therefore important. The size of the two-loop term compared to the
one-loop term is expected on general ground to be of order: 
\begin{equation}
\frac{D_{2}}{D_{1}}\,\approx \,\frac{C_{A}}{C_{F}}\,=\,2.25,
\end{equation}
since in two-loop order gluons start to be radiated by gluons istead of
quarks, the former having a larger colour charge $C_{A}$ instead of $C_{F}.$

Let us now consider the inverse Mellin transform (\ref{MellinInverse}),
releasing the frozen coupling approximation; in formula (\ref{generale})
the coupling runs over the whole integration range.

The form factor in $N$-space is computed in such a way that
infrared-pole effects appear in a sharp way for $N>N_{c},$ where it acquires
an unphysical (imaginary) part. In other terms,
the numerical distribution is not real for any value of N because of the
integration over the Landau pole.
 An exact numerical  evaluation of the inverse transform
then requires a prescription for the pole. An alternative strategy is to
give a prescription for the infrared pole directly in $N$-space, in such a
way that the form factors are well-defined for any $N.$ In general, this
results into a softening of the form factor for large $N.$ It is then not
necessary to give a prescription for the pole in the inverse transform.
 
We have used the minimal prescription (mp) \cite{mangano},
over two different paths (with the same results);
the first path was
made by two straight lines parallel to the negative real axes,
closed by a half-circle centered arond the origin and 
crossing the positive axes between the origin and the first Landau pole;
the second path was composed by two lines almost vertical,
 meeting on the positive real axes
between the origin and the first Landau pole.
The precise crossing point is irrelevant,
 as far as it is before the Landau pole.

\begin{figure}[b]
\par
\begin{center}
\epsfxsize=0.80\textwidth
\epsfysize=0.50\textheight
\leavevmode\epsffile{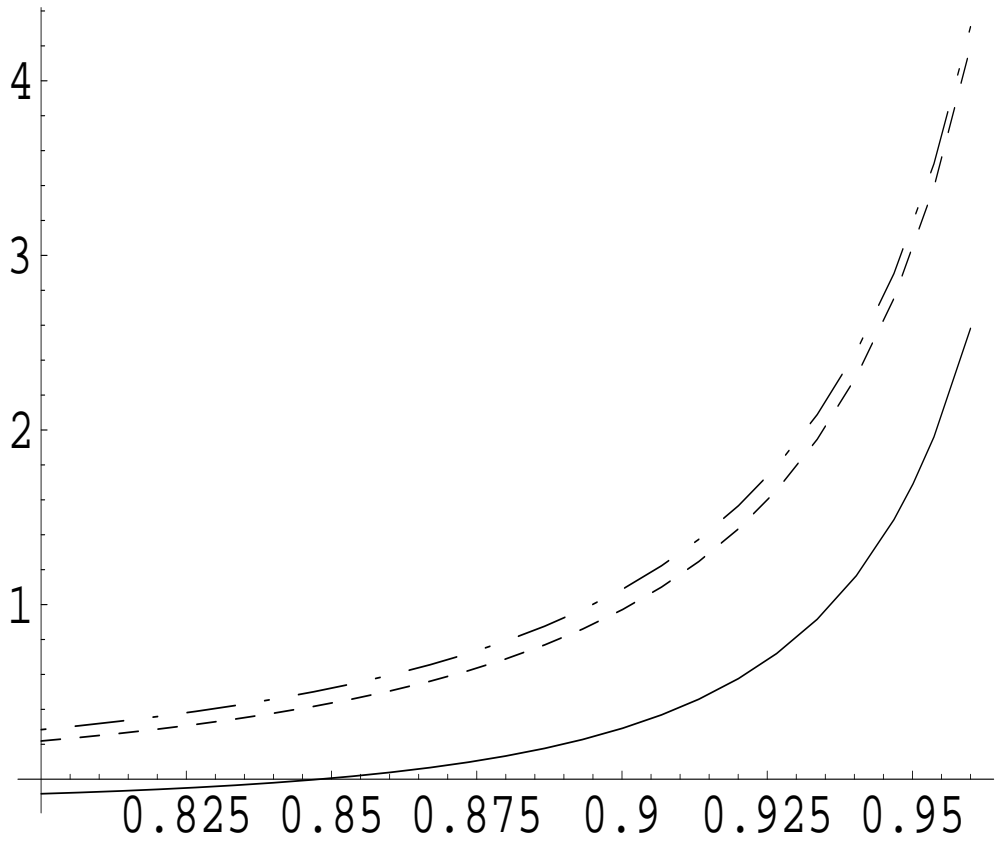}
\end{center}
\caption{{\rm{
mp numerical distribution (solid line), NLO (dashed line)
and NNLO (dot-dashed line) analytic distributions
of $f(x)$,
 at $\alpha_S=0.1$.}}}
\label{fig_finale}
\end{figure}

The inverse  Mellin transform can be also derived analytically, 
as seen in section \ref{SecInv}. 
We want to compare
 the analytic and numerical
results, as we did before in the frozen coupling approximation.
In this case, however, the Landau pole contribution plays an important
role and we 
need to reach regions of higher energy ($\alpha_S=0.1$), where 
 the first Landau pole (see eq. (\ref{firstLP}))
occurs at  larger values of $N$
($N= e^{1/ 2 \beta_0 \alpha_S} \simeq 3000-4000 $).
In these regions,
the perturbative resumming
keeps under control the infrared divergences and 
cancels the oscillatory behaviour.
We compare the analytic (NLO and NNLO) and numerical
(NNLO) plots in fig. \ref{fig_finale} (at $\alpha_S=0.1$).

\begin{center}
\bigskip Acknowledgements
\end{center}

We wish to thank S. Catani for discussions.
G.R. thanks the CERN theory department for hospitality during
 part of this work.

\end{document}